\documentclass[10pt]{bmc_article}

\usepackage{cite} 
\usepackage{url}  
\usepackage{ifthen}  
\usepackage{amsmath}
\usepackage{amssymb}
\usepackage{verbatim}
\usepackage{graphicx}
\usepackage{multicol}   
\usepackage{subfigure}
\usepackage{color}
\usepackage{hyperref}
\usepackage[utf8]{inputenc} 
\renewcommand{\d}{{\rm d}}
\newcommand{\e}{{\rm e}}
\renewcommand{\H}{\mathcal{H}}
\newcommand{\G}{\mathcal{G}}
\newcommand{\D}{\mathcal{D}}
\newcommand{\zt}{\zeta(\theta)}
\newcommand{\Df}{{\rm D}f}

\newcommand{\PD}[2]{\frac{\partial #1}{\partial #2}}
\newcommand{\FD}[2]{\frac{\d #1}{\d #2}}
\newcommand{\ut}{u(\theta)}
\DeclareMathSymbol{\Zset}{\mathalpha}{AMSb}{"5A}

\DeclareMathSymbol{\RSet}{\mathalpha}{AMSb}{"52}

\newboolean{publ}

\newenvironment{bmcformat}{\begin{raggedright}\baselineskip20pt\sloppy\setboolean{publ}{false}}{\end{raggedright}\baselineskip20pt\sloppy}

\begin{document}
\begin{bmcformat}


\title{Phase-amplitude descriptions of neural oscillator models}
 

\author{K C A Wedgwood \correspondingauthor$^1$%
       \email{K C A Wedgwood \correspondingauthor - pmxkw2@nottingham.ac.uk}%
      \and
         K K Lin $^2$%
         \email{K K Lin\ - klin@math.arizona.edu}
         \and
         R Thul $^1$
         \email{R Thul - ruediger.thul@nottingham.ac.uk}
       and 
         S Coombes$^1$%
         \email{S Coombes - stephen.coombes@nottingham.ac.uk}%
      }


\address{%
    \iid(1)School of Mathematical Sciences, University of Nottingham, Nottingham, NG7 2RD, UK.\\
    \iid(2)Department of Applied Mathematics, University of Arizona, Tucson, AZ, USA
}%

\maketitle


\begin{abstract}

Phase oscillators are a common starting point for the reduced description of many single neuron models that exhibit  a strongly attracting limit cycle.  The framework for analysing such models in response to weak perturbations is now particularly well advanced, and has allowed for the development of a theory of weakly connected neural networks.  
However, the \textit{strong-attraction} assumption may well not be the natural one for many neural oscillator models.  For example, the popular conductance based Morris-Lecar model is known to respond to periodic pulsatile stimulation in a chaotic fashion that cannot be adequately described with a phase reduction.  In this paper, we generalise the phase description that allows one to track the evolution of distance from the cycle as well as phase on cycle.  We use a classical technique from the theory of ordinary differential equations that makes use of a moving coordinate system to analyse periodic orbits.  The subsequent \textit{phase-amplitude} description is shown to be very well suited to understanding the response of the oscillator to external stimuli (which are not necessarily weak).  We consider a number of examples of neural oscillator models, ranging from planar through to high dimensional models, to illustrate the effectiveness of this approach in providing an improvement over the standard phase-reduction technique.  As an explicit application of this phase-amplitude framework, we consider in some detail the response of a generic planar model where the strong-attraction assumption does not hold, and examine the response of the system to periodic pulsatile forcing.  In addition, we explore how the presence of dynamical \textit{shear} can lead to a chaotic response.
\end{abstract}

\ifthenelse{\boolean{publ}}{\begin{multicols}{2}}{}


\section*{Keywords}
phase-amplitude, oscillator, chaos, non-weak coupling

\section{Introduction}

One only has to look at the plethora of papers and books on the topic of phase oscillators in mathematical neuroscience to see the enormous impact that this tool from dynamical systems theory has had on the way we think about describing neurons and neural networks.  Much of this work has its roots in the theory of ordinary differential equations (ODEs) and has been promoted for many years in the work of Winfree \cite{Winfree01}, Guckenheimer \cite{Guckenheimer75}, Holmes \cite{Cohen88}, Kopell \cite{Kopell86}, Ermentrout \cite{Ermentrout81} and Izhikevich \cite{Izhikevich07} to name but a few.  For a recent survey we refer the reader to the book by Ermentrout and Terman \cite{Ermentrout10}.
At heart, the classic phase reduction approach recognises that if a high dimensional nonlinear dynamical system (as a model of a neuron) exhibits a stable limit cycle attractor then trajectories near the cycle can be projected onto the cycle.  

A natural phase variable is simply the time along the cycle (from some arbitrary origin) relative to the period of oscillation.  The notion of phase can even be extended off the cycle using the concept of isochrons \cite{Winfree01}. They provide global information about the `latent phase', namely the phase that will be asymptotically returned to for a trajectory with initial data within the basin of attraction of an exponentially stable periodic orbit.  More technically, isochrons can be viewed as the leaves of the invariant foliation of the stable manifold of a periodic orbit \cite{Josic:2006}.  In rotating frame coordinates given by phase and the leaf of the isochron
foliation the system has a skew-product structure, i.e. the equation of the phase decouples. However, it is a major challenge to find the isochron foliation, and since it relies on the knowledge of the limit cycle it can only be found in special cases or numerically.  There are now a number of complementary techniques that tackle this latter challenge, and in particular we refer the reader to work of Guillamon and Huguet \cite{Guillamon2009} (using Lie symmetries) and Osinga and Moehlis \cite{Osinga10} (exploiting numerical continuation).  More recent work by Mauroy and Mezic \cite{Mauroy2012} is especially appealing as it uses a simple forward integration algorithm, as illustrated in Fig.~\ref{Fig:isochronsFourierSL} for a Stuart-Landau oscillator.  However, it is more common to side-step the need for constructing global isochrons by restricting attention to a small neighbourhood of the limit cycle, where dynamics can simply be recast in the reduced form $\dot{\theta}=1$, where $\theta$ is the phase around a cycle.  This reduction to a phase description gives a nice simple dynamical system, albeit one that cannot describe evolution of trajectories in phase-space that are far away from the limit cycle.  However, the phase reduction formalism is useful in quantifying how a system (on or close to a cycle) responds to weak forcing, via the construction of the infinitesimal phase response curve (iPRC).  For a given high dimensional conductance based model this can be solved for numerically, though for some normal form descriptions closed form solutions are also known \cite{Brown04}.  The iPRC at a point on cycle is equal to the gradient of the (isochronal) phase at that point.  This approach forms the basis for constructing models of weakly interacting oscillators, where the external forcing is pictured as a function of the phase of a firing neuron.  This has led to a great deal of work on phase-locking and central pattern generation in neural circuitry (see for example \cite{Hoppensteadt97}).  Note that the work in \cite{Guillamon2009} goes beyond the notion of iPRC and introduces infinitesimal phase response surfaces (allowing evaluation of phase advancement even when the stimulus is off cycle), and see also the work in \cite{Achuthan2009} on non-infinitesimal PRCs.
\begin{figure}[h!]
\begin{center}
\includegraphics[width=3.in]{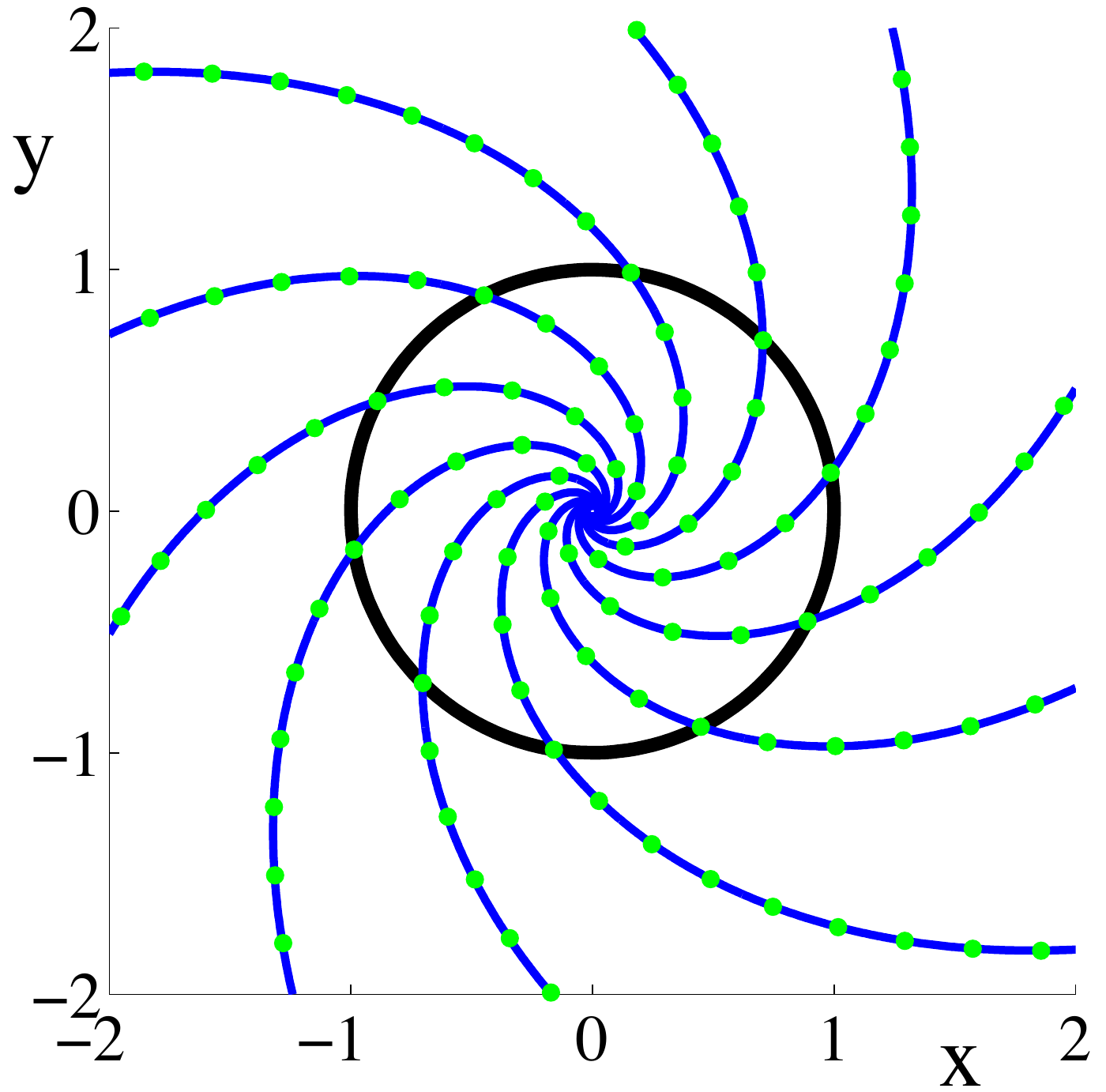} 
\caption{\label{fig:isochronsSL} Isochrons of a Stuart-Landau oscillator model:
$\dot{x} = \lambda x/2 - (\lambda c/2 + \omega)y - \lambda(x^2+y^2)(x-cy)/2$, $\dot{y} = (\lambda c/2 + \omega)x +\lambda y/2 -\lambda(x^2+y^2)(cx+y)/2$.  The black curve represents the periodic orbit of the system, which is simply the unit circle for this model. The blue curves are the isochrons obtained using the (forward) approach of Mauroy and Mezic \cite{Mauroy2012}.  The green dots are analytically obtained isochronal points \cite{Yoshimura10}. Parameter values are $\lambda=2.0$, $c=1.0$ and $\omega=1.0$. 
\label{Fig:isochronsFourierSL}
}
\end{center}
\end{figure}

The assumption that phase alone is enough to capture the essentials of neural response is one made more for mathematical convenience than being physiologically motivated.  Indeed for the popular type I Morris-Lecar (ML) firing model with standard parameters, direct numerical simulations with pulsatile forcing show responses that cannot be explained solely with a phase model \cite{Lin2012}.  The failure of a phase description is in itself no surprise and underlies why the community emphasises the use of the word \textit{weakly} in the phrase ``weakly connected neural networks".  
Indeed, there are a number of potential pitfalls when applying phase reduction techniques to a system that is not in a weakly forced regime. The typical construction of the phase response curve uses only linear information about the isochrons and nonlinear effects will come into play the further we move away from the limit cycle. This problem can be diminished by taking higher order approximations to the isochrons and using this information in the construction of a higher order PRC \cite{Demir2010}. Even using perfect information about isochrons, the phase reduction still assumes persistence of the limit-cycle and instantaneous relaxation back to cycle. However, the presence of nearby invariant phase-space structures such as (unstable) fixed point and invariant manifolds may result in trajectories spending long periods of time away from the limit cycle.  Moreover, strong forcing will necessarily take one away from the neighbourhood of a cycle where a phase description is expected to hold.  Thus developing a reduced description which captures some notion of \textit{distance from cycle} is a key component of any theory of forced limit cycle oscillators.  The development of phase-amplitude models that better characterise the response of popular high dimensional single neuron models is precisely the topic of this paper.
Given that it is a major challenge to construct an isochronal foliation we use non-isochronal phase-amplitude coordinates as a practical method for obtaining a more accurate description of neural systems.
Recently Medvedev \cite{Medvedev2010a} has used this approach to understand in more detail the synchronisation of linearly coupled stochastic limit cycle oscillators.

In section \ref{sec:phaseAmp}, we consider a general coordinate transformation which recasts the dynamics of a system in terms of phase-amplitude coordinates. This approach is directly taken from the classical theory for analysing periodic orbits of ODEs, originally considered for planar systems in \cite{Diliberto1950}, and for general systems in \cite{Hale1969}.  We advocate it here as one way to move beyond a purely phase-centric perspective.  We illustrate the transformation by applying it to a range of popular neuron models.  In section \ref{sec:forced}, we consider how inputs to the neuron are transformed under these coordinate transformations and derive the evolution equations for the forced phase-amplitude system.  This reduces to the standard phase description in the appropriate limit.  Importantly, we show that the behaviour of the phase-amplitude system is much more able to capture that of the original single neuron model from which it is derived.  Focusing on pulsatile forcing we explore the conditions for neural oscillator models to exhibit shear induced chaos \cite{Lin2012}. Finally in section \ref{sec:discussion}, we discuss the relevance of this work to developing a theory of network dynamics that can improve upon the standard weak coupling approach. 

\section{Phase-amplitude coordinates}
\label{sec:phaseAmp}

Throughout this paper, we study the dynamics prescribed by the system $\dot{x}=f(x)$, $x \in \RSet^n$, with solutions $x=x(t)$ that satisfy $x(0)=x_0 \in \RSet^n$.  We will assume that the system admits an attracting hyperbolic periodic orbit (namely one zero Floquet exponent and the others having negative real part), with period $\Delta$, such that $u(t)=u(t+\Delta)$.  A phase $\theta \in [0,\Delta)$ is naturally defined from $\theta(u(t))=t \! \mod \Delta$.  It has long been known in the dynamical systems community how to construct a coordinate system based on this notion of phase as well as a \textit{distance} from cycle, see \cite{Hale1969} for a discussion.  In fact, Ermentrout and Kopell \cite{Ermentrout90} made good use of this approach to derive the phase-interaction function for networks of weakly connected limit-cycle oscillators in the limit of infinitely fast attraction to cycle.  However, this assumption is particularly extreme and unlikely to hold for a broad class of single neuron models.  Thus, it is interesting to return to the full phase-amplitude description.  In essence, the transformation to these coordinates involves setting up a {\it moving orthonormal system} around the limit cycle.  One axis of this system is chosen to be in the direction of the tangent vector along the orbit, and the remaining are chosen to be orthogonal.  We introduce the normalised tangent vector $\xi$ as
\begin{equation}
\xi(\theta) = \FD{u}{\theta}\Big / \left|\FD{u}{\theta}\right|.
\label{xi}
\end{equation} 
The remaining coordinate axes are conveniently grouped together as the columns of an $n\times (n-1)$ matrix $\zeta$.  In this case we can write an arbitrary point $x$ as
\begin{equation}
	x(\theta,\rho)=u(\theta)+\zeta(\theta)\rho .
\label{eq:trans}
\end{equation} 
Here, $|\rho|$ represents the Euclidean distance from the limit cycle. A caricature, in $\RSet^2$, of the coordinate system along an orbit segment is shown in Fig.~\ref{fig:PAcoords}. Through the use of the variable $\rho$, we can consider points away from the periodic orbit. Rather than being isochronal, lines of constant $\theta$ are simply straight lines that emanating from a point on the orbit in the direction of the normal. The technical details of specifying the orthonormal coordinates forming $\zeta$ are discussed in Appendix~\ref{app:trans}.

\begin{figure}[h!]
\begin{center}
\includegraphics[width=3.in]{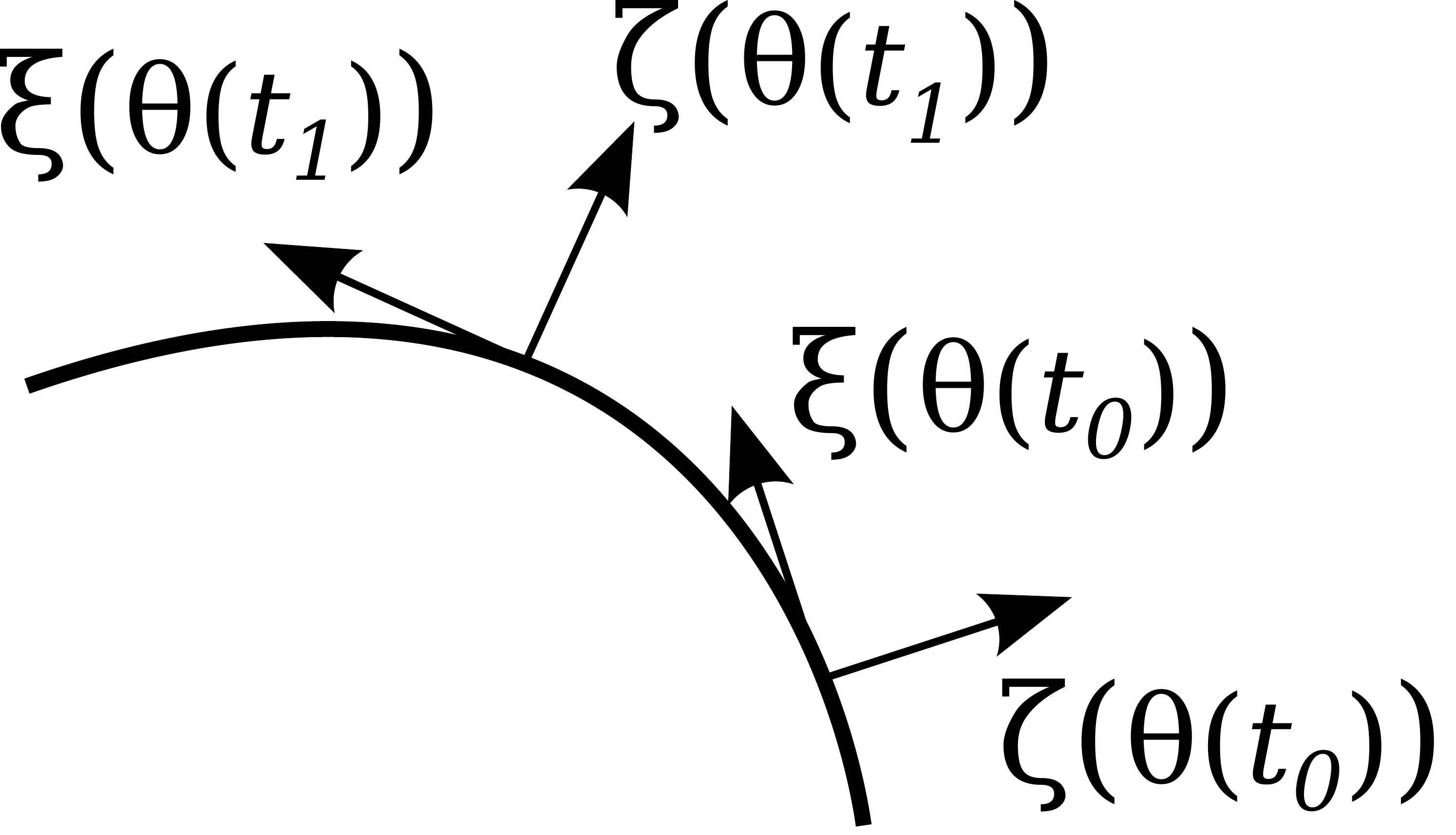} 
\caption{ \label{fig:PAcoords} Demonstration of the moving orthonormal coordinate system along an orbit segment. As $t$ evolves from $t_0$ to $t_1$, the coordinates vary smoothly. In this planar example, $\zeta$ always points to the outside of the orbit. }
\end{center}
\end{figure}

Upon projecting the dynamics onto the moving orthonormal system, we obtain the dynamics of the transformed system:
\begin{align}
	\dot{\theta}=1+f_1(\theta,\rho) , \quad  \dot{\rho}=A(\theta)\rho+f_2(\theta,\rho) .
 	\label{eq:main2}
\end{align}
where
\begin{align}
	f_1(\theta,\rho)&=-h^T(\theta,\rho)\FD{\zeta}{\theta}\rho+h^T(\theta,\rho)\left[f(u+\zeta\rho)-f(u)\right] , \\
	f_2(\theta,\rho)&=-\zeta^T\FD{\zeta}{\theta}\rho f_1+\zeta^T\left[f(u+\zeta\rho)-f(u)-\Df\zeta\rho\right] ,
\end{align}
\begin{align}
	h(\theta,\rho) = \left[\left| \FD{u}{\theta}\right| + \xi^T \FD{\zeta}{\theta}\rho \right]^{-1} \xi , \quad A(\theta)=\zeta^T\left[-\FD{\zeta}{\theta}+\Df\zeta\right] ,
\end{align}
and $\D f$ is the Jacobian of the vector field $f$, evaluated along the periodic orbit $u$.
The derivation of this system may be found in appendix \ref{app:trans}. It is straight-forward to show that 	$f_1(\theta,\rho) \rightarrow 0 \mbox{ as } |\rho |\rightarrow 0$, $f_2(\theta,0) = 0$ and that $\partial f_2 (\theta,0)/\partial\rho = 0$. 
In the above, $f_1$ captures the {\it shear} present in the system, that is, whether the speed of $\theta$ increases or decreases dependent on the distance from cycle. A precise definition for shear is given in \cite{Ott2010}.
Additionally, $A(\theta)$ describes the $\theta$-dependent rate of attraction or repulsion from cycle.  

\begin{figure}
\begin{center}
\includegraphics[width=4.in]{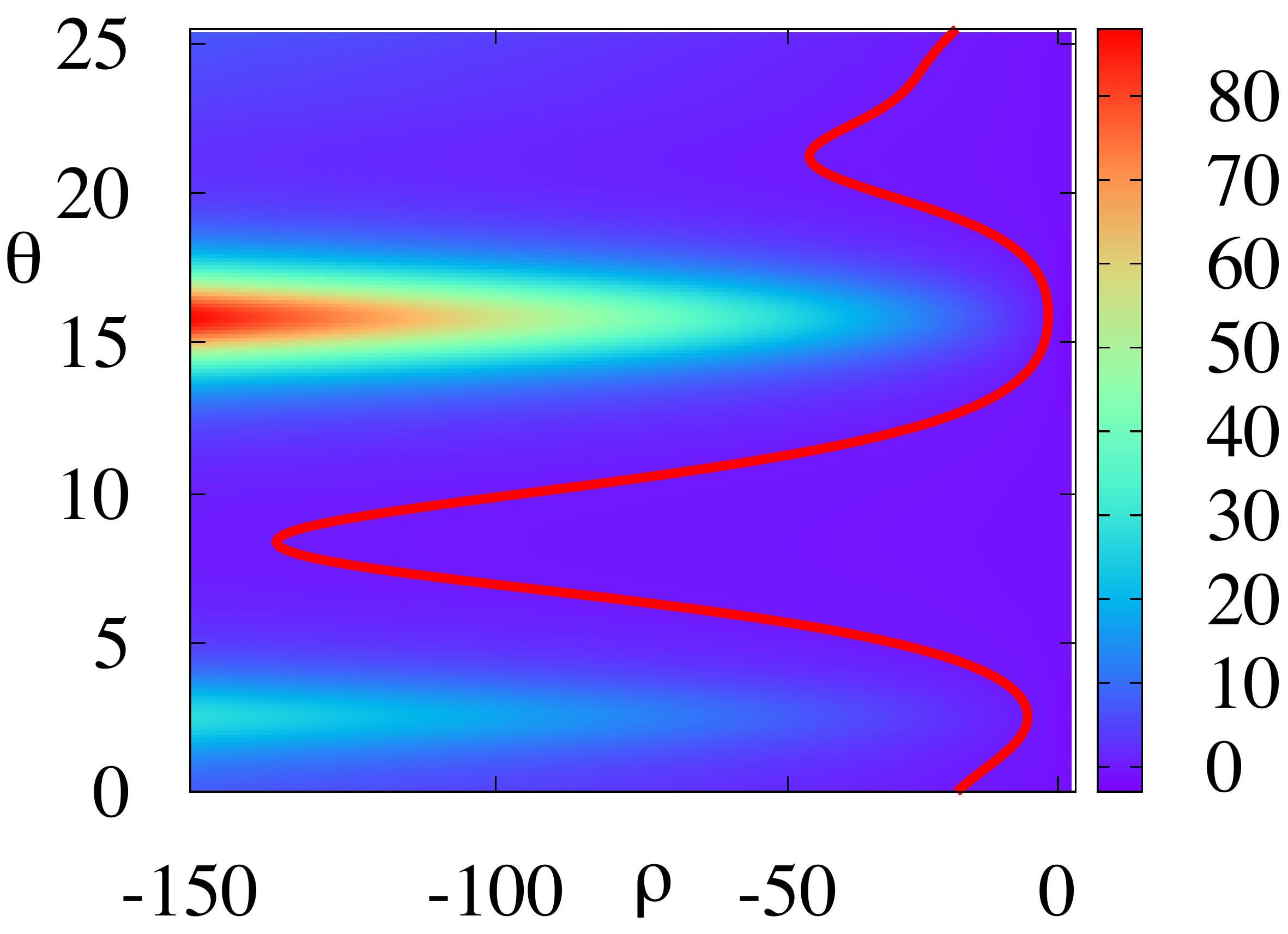} 
\caption{ \label{fig:MLdetK} This figure shows the determinant of the phase-amplitude transformation for the ML model. Colours indicate the value of $K$. The red contour indicates where $K = 0$, and thus where the coordinate transformation breaks down. Note how all of the values for which this occurs have $\rho<0$. Parameter values as in Appendix \ref{ssec:ML}.}
\end{center}
\end{figure}
It is pertinent to consider where this coordinate transformation breaks down, that is, where the determinant of the Jacobian of the transformation $K=\det 
[\partial x /\partial \theta ~\partial x/\partial \rho ]$ vanishes.  This never vanishes on-cycle (where $\rho=0$), but may do so for some $|\rho|= k>0$. This sets an upper bound on how far away from the limit cycle we can describe the system using these phase-amplitude coordinates. 
In Fig.~\ref{fig:MLdetK} we plot the curve along which the transformation breaks down for the ML model.  We observe that, for some values of $\theta,k$ is relatively smaller. The breakdown occurs where lines of constant $\theta$ cross and thus the transformation ceases to be invertible, and these values of $\theta$ correspond to points along which the orbit has high curvature. We note that this issue is less problematical in higher dimensional models.

If we now consider the driven system 
\begin{equation}
	\dot{x} = f(x) + \varepsilon g(x,t) , \quad x\in\RSet^n ,
	\label{driven}
\end{equation}
where $\varepsilon$ is not necessarily small, we may apply the same transformation as above to obtain the dynamics in $(\theta,\rho)$ coordinates, where $\theta\in[0,\Delta)$ and $\rho\in\RSet^{n-1}$, as
\begin{align}
	\dot{\theta}&=1+f_1(\theta,\rho)+\varepsilon h^T(\theta,\rho)g(u(\theta)+\zeta(\theta)\rho,t),  \label{drivetheta}\\
	\dot{\rho}&=A(\theta)\rho+f_2(\theta,\rho)+\varepsilon\zeta^T B(\theta,\rho) g(u(\theta)+\zeta(\theta)\rho,t), \label{driverho}
\end{align}
with
\begin{equation}
	B(\theta,\rho) = \mathrm{I}_n-\FD{\zeta}{\theta}\rho h^T(\theta,\rho) ,
\end{equation}
and $\mathrm{I}_n$ is the $n\times n$ identity matrix. Here, $h$ and $B$ describe the effect in terms of $\theta$ and $\rho$ that the perturbations have. Details of the derivation are given in Appendix \ref{app:trans}. For planar models, $B=\mathrm{I}_2$.  To demonstrate the application of the above coordinate transformation, we now consider some popular single neuron models.

\subsection{A 2D conductance based model}

The ML model was originally developed to describe the voltage dynamics of barnacle giant muscle fiber \cite{Morris1981}, and is now a popular modelling choice in computational neuroscience \cite{Ermentrout10}.  It is written as a pair of coupled nonlinear ODEs of the form
\begin{equation}
    C \dot{v} = I(t) - g_{\rm l}(v-v_{\rm l}) - g_{\rm k} w(v-v_k) - g_{\rm ca}m_\infty(v)(v-v_{\rm ca}) ,\qquad
    \dot{w} = \phi\big(w_\infty(v) - w\big)/\tau_w(v).
  \label{eq:ml}
\end{equation}
Here, $v$ is the membrane voltage, whilst $w$ is a {\em gating variable}, describing the fraction of membrane ion channels that are open at any time. The first equation expresses Kirchoff's current law across the cell membrane, with $I(t)$ representing a stimulus in the form of an injected current. The detailed form of the model is completed in Appendix \ref{ssec:ML}.
The ML model has a very rich bifurcation structure.  Roughly speaking, by varying a constant current $I(t)\equiv I_0$~, one observes, in different parameter regions, dynamical regimes corresponding to sinks, limit cycles, and Hopf, saddle-node and homoclinic bifurcations, as well as combinations of the above.  These scenarios are discussed in detail in \cite{Ermentrout10} and \cite{Rinzel1989}.

As the ML model is planar, $\rho$ is a scalar, as are the functions $A$ and $f_{1,2}$. This allows us to use the moving coordinate system to clearly visualise parts of phase space where trajectories are attracted towards the limit cycle, and parts in which they move away from it, as illustrated in Fig.~\ref{fig:psolML}.  The functions $f_{1,2}$ and $A$, evaluated at $\rho=-0.1$ are shown in Fig.~\ref{fig:MLfuncEvals}. The evolution of $\theta$ is mostly constant, however we clearly observe portions of the trajectories where this is slowed, along which, $\dot{\rho}\approx 0$. In fact, this corresponds to where trajectories pass near to the saddle node, and the dynamics stall. This occurs around $\theta=17$, and in Fig.~\ref{fig:MLfuncEvals} we see that both $A(\theta)$ and $f_1(\theta,\rho)$ are indeed close to 0. The reduced velocities of trajectories here highlights the importance of considering other phase space structures in forced systems, the details of which are missed in standard phase only models. Forcing in the presence of such structures may give rise to complex and even chaotic behaviours, as we shall see in section \ref{sec:forced}.

\begin{figure}
\begin{center}
\includegraphics[width=6.in]{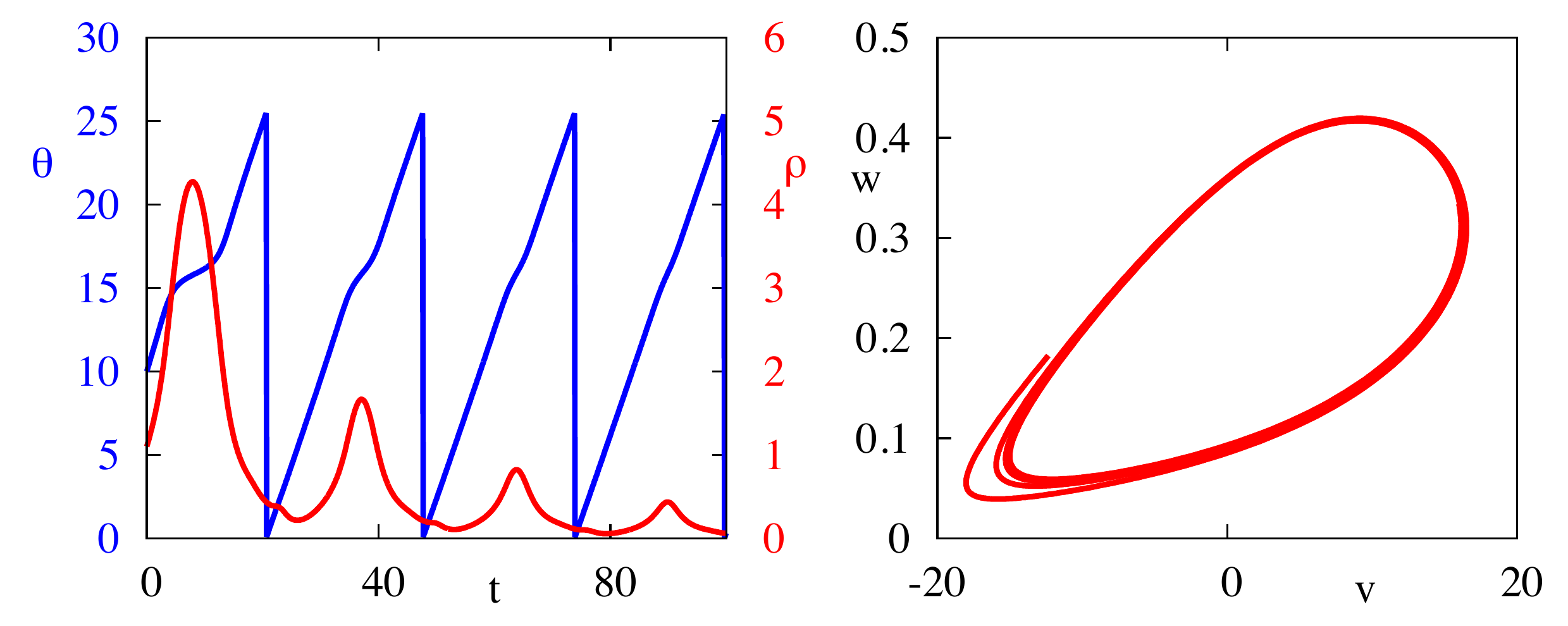} 
\caption{ \label{fig:psolML} Typical trajectory of the ML model of the transformed system. Left: Time evolution of $\theta$ and $\rho$. Right: Trajectory plotted in the $(v,w)$ phase plane. We see that when $\rho$ has a local maximum, the evolution of $\theta$ slows down,  corresponding to where trajectories pass near to the saddle node.  Parameter values as in Appendix \ref{ssec:ML}.}
\end{center}
\end{figure}
\begin{figure}[h!]
\begin{center}
\includegraphics[width=6.in]{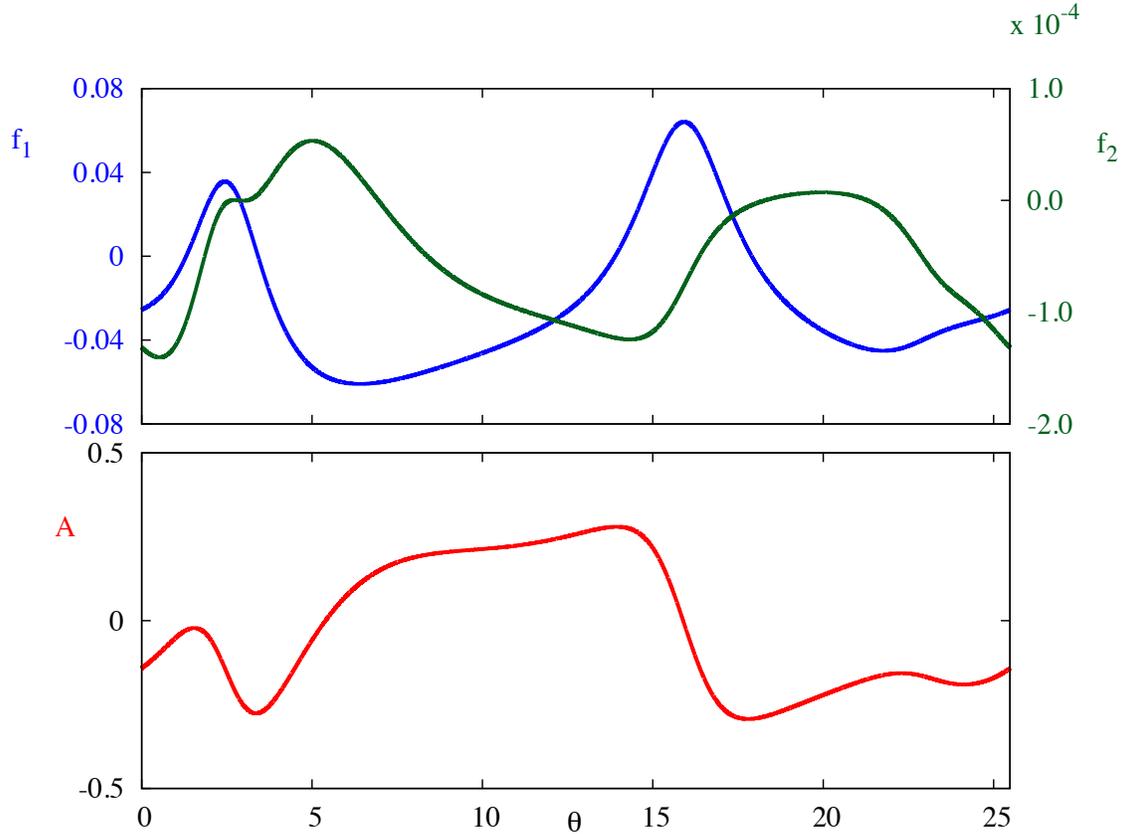} 
\caption{ \label{fig:MLfuncEvals} $f_1,f_2$ and $A$ for the ML model, evaluated at $\rho=-0.1$.  We clearly see the difference in the order of magnitude between $f_1$ and $f_2$ for small $\rho$.  Note that, although the average of $A$ over one period is negative, it is positive for a nontrivial interval of $\theta$. This corresponds to movement close to the stable manifold of the saddle node. Parameter values as in Appendix \ref{ssec:ML}.}
\end{center}
\end{figure}

In the next example, we show how the same ideas go across to higher dimensional models.

\subsection{A 4D conductance based model}

The Connor-Stevens (CS) model \cite{Connor1971} is built upon the Hodgkin-Huxley formalism and comprises a fast Na$^+$ current, a delayed K$^+$ current, a leak current and a transient K$^+$ current, termed the A-current.  The full CS model consists of 6 equations: the membrane potential, the original Hodgkin-Huxley gating variables and an activating and inactivating gating variable for the A-current. Using the method of equivalent potentials \cite{Kepler1992}, we may reduce the dimensionality of the system, to include only 4 variables. The reduced system is: 
\begin{align}
	 C \dot{v} = - F(v,u,a,b)  + I , \quad \dot{u} &= G(v,u) , \quad \dot{X} = \frac{X_\infty(v)-X}{\tau_X(v)} , \quad X \in \{a,b\} ,
\end{align}
where
\begin{equation}
F(v,u,a,b) = g_{\rm K} n_\infty^4(u)(v-v_{\rm K})+g_{\rm Na}h_\infty(u)m_\infty^3(v)(v-v_{\rm Na})+g_{\rm L}l(v-v_{\rm l})+g_{\rm a} a^3b(v-v_{\rm a}) . 
\label{eq:CS} 
\end{equation}
The details of the reduced CS model are completed in Appendix \ref{ssec:CS}. 
The solutions to the reduced CS model under the coordinate transformation may be seen in Fig.~\ref{fig:CSPA}, whilst, in Fig.~\ref{fig:CS3DPA}, we show how this solution looks in the original coordinates. As for the ML model, $\theta$ evolves approximately constantly throughout, though this evolution is sped up close to $\theta=\Delta$. The trajectories of the vector $\rho$ are more complicated, but note that there is regularity in the pattern exhibited, and that this occurs with approximately the same period as the underlying limit cycle. The damping of the amplitude of oscillations in $\rho$ over successive periods represents the overall attraction to the limit cycle, whilst the regular behaviour of $\rho$ represents the specific relaxation to cycle as shown in Fig.~\ref{fig:CS3DPA}. Additional file 1 shows a movie of the trajectory in $(v,u,b)$ coordinates with the moving orthonormal system superimposed, as well as the solution in $\rho$ for comparison.

\begin{figure}[h!]
\begin{center}
\includegraphics[width=6.in]{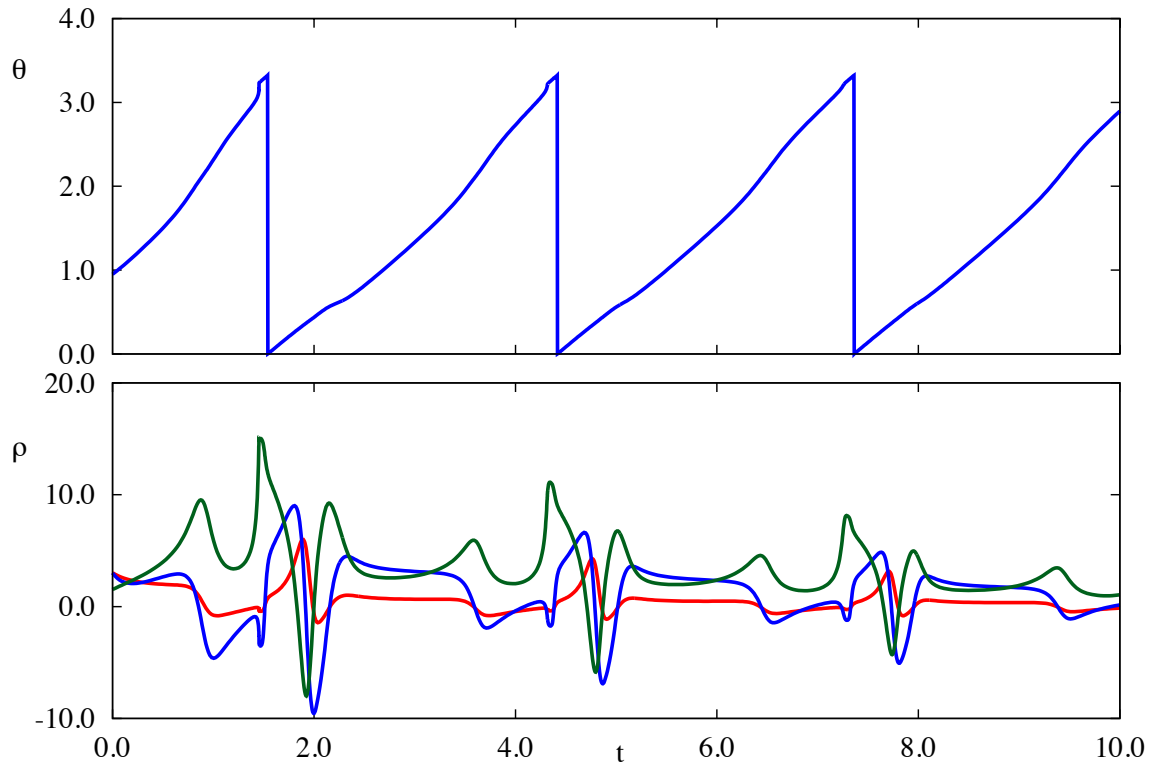} 
\caption{ \label{fig:CSPA} Solution of the transformed CS system. Top: Time evolution of $\theta$. Bottom: Time evolution of $\rho$ coordinates. Upon transforming these coordinates back to the original ones, we arrive at Fig.~\ref{fig:CS3DPA}. Parameter values given in Appendix \ref{ssec:CS}. In this parameter regime the model exhibits type I firing dynamics.}
\end{center} 
\end{figure}
\begin{figure}[h!]
\begin{center}
\includegraphics[width=6.in]{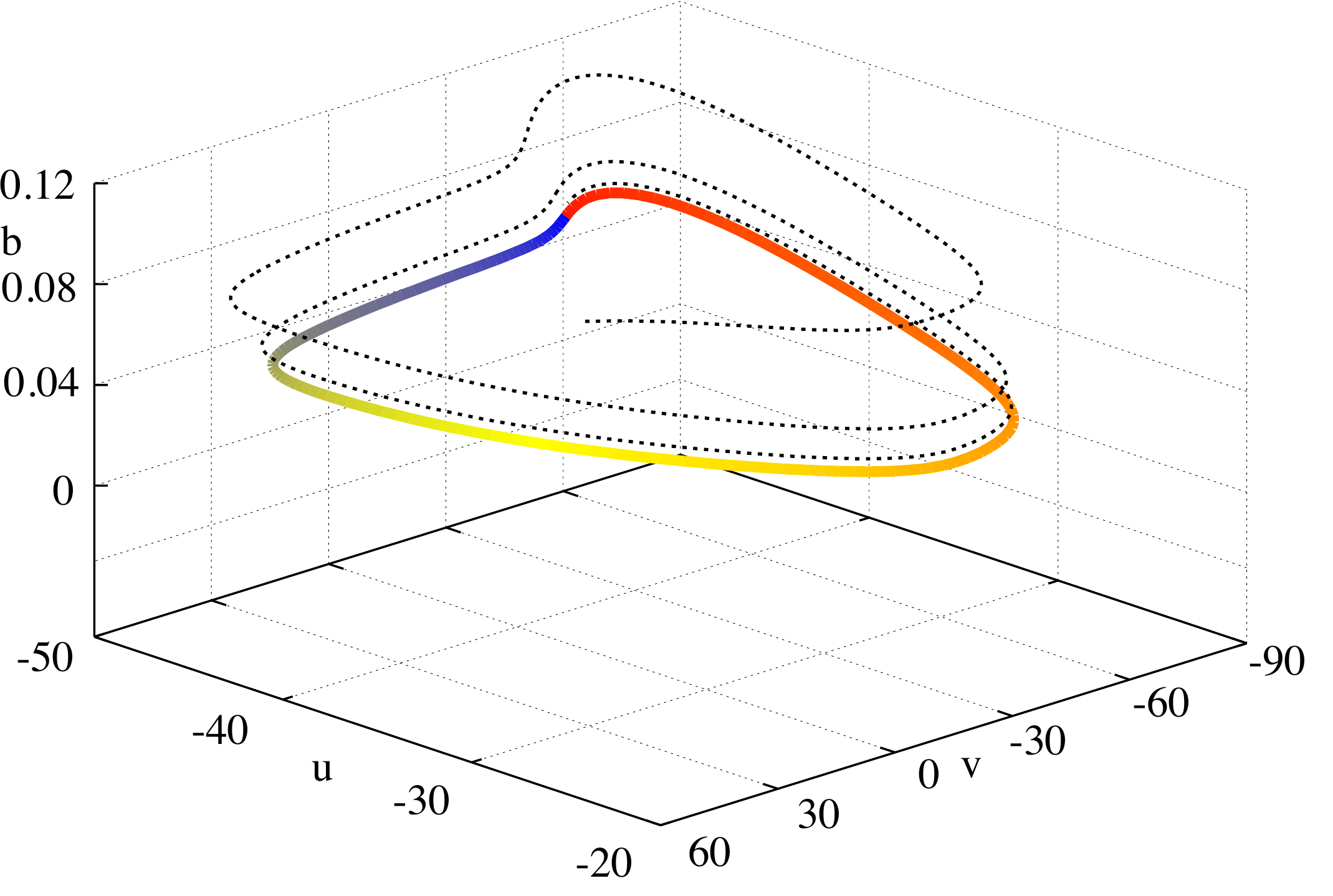} 
\caption{ \label{fig:CS3DPA} Transformed trajectory in $(v,u,b)$ space of the phase-amplitude description of the reduced CS model. The dotted black line is the phase amplitude solution transformed in the original coordinates, whilst the coloured orbit is the underlying periodic orbit, where the colour corresponds to the phase along the orbit. The solution of the phase-amplitude description of the model in $(\theta,\rho)$ coordinates is shown in Fig.~\ref{fig:CSPA}. }
\end{center}
\end{figure}

\section{Pulsatile forcing of phase-amplitude oscillators}
\label{sec:forced}

We now consider a system with time-dependent forcing, given by (\ref{driven}) with
\begin{equation}
g(x,t) = \sum_{n \in \Zset} \left( \delta(t-nT) , 0 , \dots , 0 \right)^T ,
	\label{eq:generalSystem}
\end{equation}
where $\delta$ is the Dirac $\delta$-function. This describes $T$-periodic {\it kicks} to the voltage variable. Even such a simple forcing paradigm can give rise to rich dynamics \cite{Lin2012}. For the periodically kicked ML model, shear forces can lead to chaotic dynamics as folds and horseshoes accumulate under the forcing. This means that the response of the neuron is extremely sensitive to the initial phase when the kicks occur. In terms of neural response, this means that the neuron is unreliable \cite{Lin2008}. 

The behaviour of oscillators under such periodic pulsatile forcing is the subject of a number of studies, see, e.g., \cite{Wang2001,Wang2002,Wang2003,Lin2008}.
Of particular relevance here is \cite{Lin2008}, in which a
qualitative reasoning of the mechanisms that bring about shear in such
models is supplemented by direct numerical simulations to detect the
presence of chaotic solutions. For the ML model in a parameter region
close to the homoclinic regime, kicks can cause trajectories to pass
near the saddle-node, and folds may occur as a result \cite{Lin2012}.

We here would like to compare full planar neural models to the simple model, studied in \cite{Lin2008}:
\begin{equation}
	\dot{\theta} = 1 + \sigma\rho, \qquad
	\dot{\rho} = -\lambda\rho + \varepsilon P(\theta) \sum_{n \in \Zset} \delta (t-nT) .
\label{eq:LinYoung}
\end{equation}
This system exhibits dynamical shear which, under certain conditions, can lead to chaotic dynamics. The shear parameter $\sigma$ dictates how much trajectories are `sped up' or `slowed down' dependent on their distance from the limit cycle, whilst $\lambda$ is the rate of attraction back to the limit cycle, which is independent of $\theta$. Supposing that the function $P$ is smooth but non-constant, trajectories will be taken a variable distance from the cycle upon the application of the kick. When kicks are repeated, this geometric
mechanism can lead to repeated stretching and folding of phase space.
It is clear that the larger $\sigma$ is in (\ref{eq:LinYoung}), the more
shear is present, and the more likely we are to observe the folding
effect. In a similar way, smaller values of $\lambda$ mean that the
shear has longer to act upon trajectories and again result in a greater
likelihood of chaos.  Finally, to observe chaotic response, we must
ensure that the shear forces have sufficient time to act, meaning that
$T$, the time between kicks must not be too small.

This stretching and folding action can clearly lead to the
  formation of Smale horseshoes, which are well known to lead to a type
  of chaotic behaviour.  However, horseshoes may coexist with sinks,
  meaning the resulting chaotic dynamics would be transient.  Wang and
  Young proved that under appropriate conditions, there is a set of $T$
  of positive Lebesgue measure for which the system experiences a
  stronger form of sustained, chaotic behaviour, characterised by the
  existence of a positive Lyapunov exponent for almost all initial
  conditions and the existence of a ``strange attractor''; see, e.g.,
\cite{Wang2001,Wang2002,Wang2003}.

\begin{figure}[h!]
\begin{center}
\includegraphics[width=5.in]{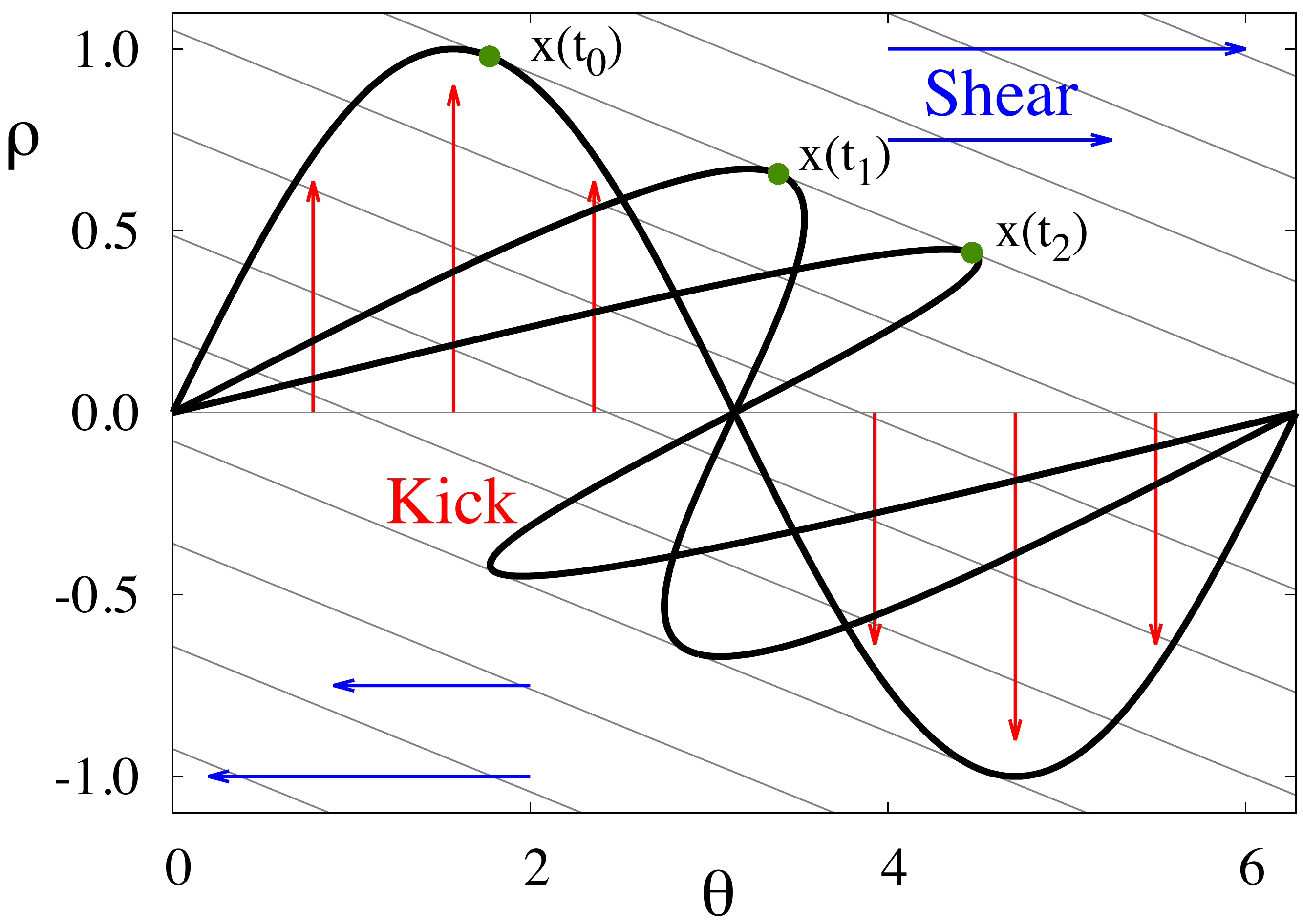} 
\caption{ \label{fig:shear} Stretch-and-fold action of a kick followed by relaxation in the presence of shear. The thin black lines are the isochrons of the system, which in the case of the linear model \eqref{eq:LinYoung}, are simply straight lines with slope $-\lambda/\sigma$. The thin gray line at $\rho=0$ represents the limit cycle, which is kicked, at $t=t_0$ by $P(\theta)=\sin(\theta)$ with strength $\varepsilon=1$ to the solid curve. After this, the orbits are allowed to evolve under the flow generated by the continuous part of the system. The dashed and dotted curves represent the image of the kicked solid curve under this flow, at times $t_1$ and $t_2$ respectively. The green marker shows how one point, $x(t_0)$ evolves under the flow, first to $x(t_1)$ and then to $x(t_2)$, following the isochron as it relaxes back to the limit cycle. The effect of the shear forces and the subsequent folding, caricatured by the blue arrows can clearly be seen. }
\end{center}
\end{figure}

By comparing with the phase-amplitude dynamics described by equations (\ref{drivetheta})-(\ref{driverho}), we see that the model of shear considered in (\ref{eq:LinYoung}) is a proxy for a more general system, with $f_1(\theta,\rho) \rightarrow \sigma \rho$, $A(\theta) \rightarrow -\lambda $ and $h(\theta,\rho) \rightarrow 0$, and $\zeta(\theta) \rightarrow P(\theta)$.

To gain a deeper insight into the phenomenon of shear induced chaos, it is pertinent to study the isochrons of the limit cycle for the linear model (\ref{eq:LinYoung}), where the isochrons are simply lines with slope $-\lambda/\sigma$. In Fig.~\ref{fig:shear}, we depict the isochrons and stretch and fold action of shear. The bold thin gray line at $\rho=0$ is kicked, at $t=t_0$, to the bold solid curve, where $P(\theta)=\sin(\theta)$, as studied in \cite{Lin2012} and this curve is allowed to evolve under the dynamics with no further kicks through the dashed curve at $t=t_1$ and ultimately to the dotted curve at $t=t_2$, which may be considered as evolutions of the solid curve by integer multiples of the natural period of the oscillator. Every point of the dotted curve traverses the isochron it is on at $t_0$ until $t_2$. The green marker shows an example of one such point evolving along its associated isochron.  The folding effect of this is clear in the figure, and further illustrated in the video in Additional file 2.

This simple model with a harmonic form for $P(\theta)$ provides insight into how strange attractors can be formed. Kicks along the isochrons or ones that map isochrons to one another will not produce strange attractors, but merely phase-shifts. What causes the stretching and folding is the variation in how far points are moved as measured in the direction transverse to the isochrons. For the linear system \eqref{eq:LinYoung} variation in this sense is generated by any non-constant $P(\theta)$; the larger the ratio $\sigma\varepsilon/\lambda$, the larger the variation (see \cite{Lin2012} for a recent discussion).  

The formation of chaos in the ML model is shown in Fig.~\ref{fig:accumulateFoldsML}.  Here, we plot the response to periodic pulsatile forcing, given by (\ref{eq:generalSystem}), in the $(v,w)$ coordinate system.  This clearly illustrates a folding of phase space around the limit-cycle, and is further portrayed in the video in Additional file 3.
We now show how this can be understood using phase-amplitude coordinates.

\begin{figure}[h!]
\begin{center}
\includegraphics[width=6.in]{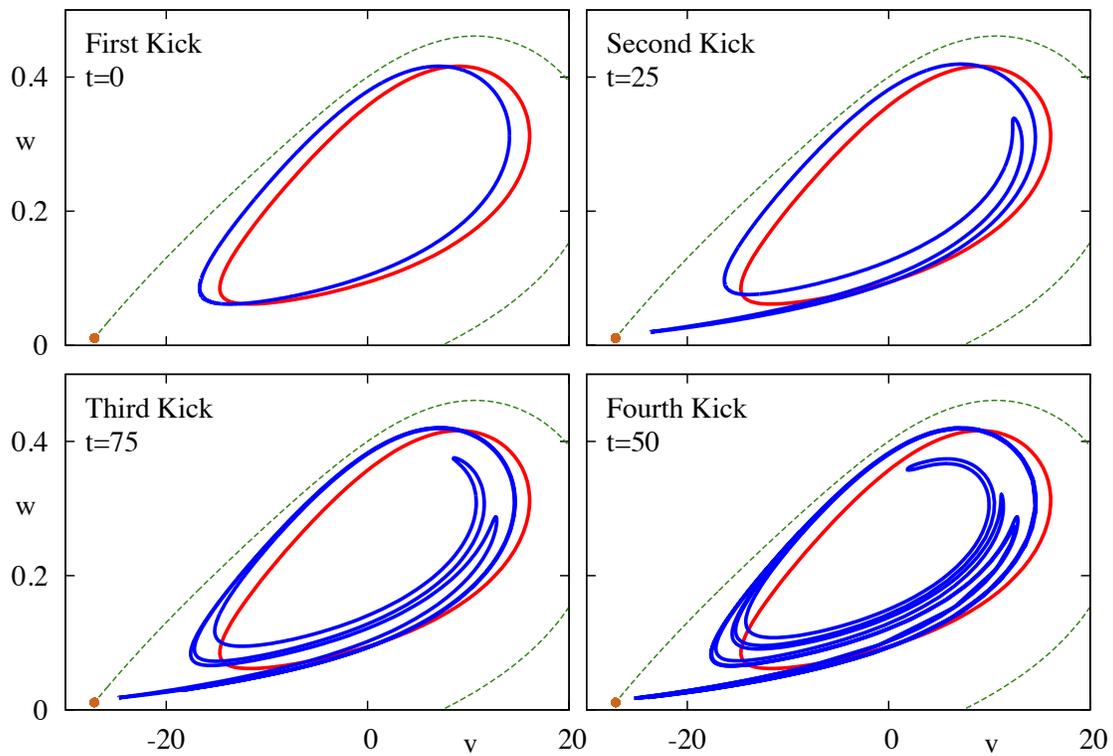} 
\caption{ \label{fig:accumulateFoldsML} Shear induced folding in the ML model, with parameters as in Fig.~\ref{fig:MLfuncEvals}. The red curve in all panels represents the limit cycle of the unperturbed system, whilst the green dotted line represents the stable manifold of the saddle node, indicated by the orange marker. We begin distributing points along the limit cycle and then apply an instantaneous kick taking $v\mapsto v+\varepsilon$, where $\varepsilon=-2.0$, leaving $w$ unchanged. This essentially moves all phase points to the left, to the blue curve. The successive panels show the image of this set of points after letting points evolve freely under the system defined by the ML equations, and then apply the kick again. The curves shown are the images of the initial phase points just after each kick, as indicated in the figure. We can clearly observe the shear induced folding.  Parameter values as in Appendix \ref{ssec:ML}.}
\end{center}
\end{figure}

Compared to the phenomenological system \eqref{eq:LinYoung}, models written in phase-amplitude coordinates as (\ref{drivetheta}-\ref{driverho}) have two main differences.
The intrinsic dynamics (without kicks) are nonlinear, and the kick terms appear in both equations for $\dot{\theta}$ and $\dot{\rho}$ (not just $\dot{\rho}$).  Simulations of (\ref{drivetheta})-(\ref{driverho}) for both the FHN and ML models, with $\varepsilon=0.1$, show that the replacement of $f_1(\theta,\rho)$ by $\sigma \rho$, dropping $f_2(\theta,\rho)$ (which is quadratic in $\rho)$, and setting $A(\theta)=-\lambda$ does not lead to any significant qualitative change in behaviour (for a wide range of $\sigma, \lambda >0$).  We therefore conclude that, at least when the kick amplitude $\varepsilon$ is not too large, it is more important to focus on the form of the forcing in the phase-amplitude coordinates. In what follows, we are interested in discovering the effects of different functional forms of the forcing term multiplying the $\delta$-function, keeping other factors fixed. As examples, we choose those forcing terms given by transforming the FHN and the ML models into phase-amplitude coordinates. To find these functions, we first find the attracting limit cycle solution in the ML model \eqref{eq:ml} and FHN model \eqref{eq:FHN} using a periodic boundary value problem solver and set up the orthonormal coordinate system around this limit cycle. Once the coordinate system is established, we evaluate the functions $h(\theta,\rho)$ and $B(\theta,\rho)$ (that appear in equations (\ref{drivetheta}) and (\ref{driverho})). For planar systems we have simply that $B(\theta,\rho)=\mathrm{I}_2$. Using the forcing term \eqref{eq:generalSystem}, we are only considering perturbations to the voltage component of our system and thus only the first component of $h$ and $B$ will make a nontrivial contribution to the dynamics.  We define $P_1$ as the first component of $h$ and $P_2$ as the first component of $\zeta$. We wish to force each system at the same ratio of the natural frequency of the underlying periodic orbit. To ease comparison between the system with the ML forcing terms and the FHN forcing terms, we rescale $\theta\mapsto\theta/\Delta$ so that $\theta\in[0,1)$ in what follows. Implementing the above choices leads to 
\begin{equation}
	\dot{\theta} = 1 + \sigma\rho + \varepsilon P_1(\theta,\rho) \sum_{n \in \Zset} \delta(t-nT) , \qquad
	\dot{\rho} = -\lambda\rho + \varepsilon P_2(\theta) \sum_{n \in \Zset} \delta(t-nT) \label{ab}.
\end{equation}
It is important to emphasise that $P_{1,2}$ are determined by the underlying single neuron model (unlike in the toy model (\ref{eq:LinYoung})). As emphasised in \cite{Catlla2008}, one must take care in the treatment of the state dependent `jumps' caused by the $\delta$-functions in (\ref{ab}) to accommodate the discontinuous nature of $\theta$ and $\rho$ at the time of the kick. To solve (\ref{ab}), we approximate $\delta(t)$ with a normalised square pulse $\delta_\tau(t)$ of the form
\begin{equation}
	\delta_\tau(t) = 
	\begin{cases}
		0 & t\leq 0 , \\
		1/\tau & 0< t \leq \tau , \\
		0 & t>\tau ,\\
	\end{cases}
	\label{deltaapprox}
\end{equation}
where $\tau\ll1$. This means that for $(n-1)T+\tau < t \leq nT$, $n\in\mathbb{Z}$, the dynamics are governed by the linear system $(\dot{\theta},\dot{\rho})=(1 + \sigma \rho,-\lambda \rho)$.  This can be integrated to obtain the state of the system just before the arrival of the next kick, $(\theta_n,\rho_n) \equiv (\theta(nT),\rho(nT))$, in the form
\begin{align}
	\theta_{n} & = \left[\theta(t) + nT-t + \frac{\sigma}{\lambda}\rho(t) \left(1-\mathrm{e}^{-\lambda(nT-t)}\right) \right]\!\!\!\!  \mod 1  \label{eq:linear1} , \\
	\rho_{n} & = \rho(t) \mathrm{e}^{-\lambda(nT-t)} \label{eq:linear2} .
\end{align}
In the interval $nT<t < nT+\tau$ and using (\ref{deltaapprox}) we now need to solve the system of nonlinear ODEs
\begin{equation}
	\dot{\theta}  = 1 + \sigma \rho + \frac{\varepsilon}{\tau} P_1(\theta,\rho) , \qquad
	\dot{\rho}  = -\lambda \rho + \frac{\varepsilon}{\tau}P_2(\theta)  .
\end{equation}
Rescaling time as $t=nT+\tau s$, with $0 \leq s \leq 1$, and writing the solution $(\theta,\rho)$ as a regular perturbation expansion in powers of $\tau$ as $(\theta(s), \rho(s)) = (\theta_0(s) + \tau  \theta_1(s),  \rho_0(s) + \tau  \rho_1(s)) + \ldots ~$, we find after collecting terms of leading order in $\tau$ that the pair $(\theta_0(s),\rho_0(s))$ is governed by
\begin{equation}
	\FD{\theta_0}{s}  =  \varepsilon P_1(\theta_0(s), \rho_0(s)) ,\qquad
	\FD{\rho_0}{s}  = \varepsilon P_2(\theta_0(s)),  \qquad 0 \leq s \leq 1 \label{eq:kick} ,
\end{equation}
with initial conditions $(\theta_0(0),\rho_0(0))=(\theta_{n},\rho_{n})$. 
The solution $(\theta_0(1),\rho_0(1))\equiv (\theta_{n}^+,\rho_{n}^+)$ (obtained numerically) can then be taken as initial data $(\theta(t),\rho(t))=(\theta_{n}^+,\rho_{n}^+)$ in (\ref{eq:linear1})-(\ref{eq:linear2}) to form the stroboscopic map
\begin{align}
	\theta_{n+1} & = \left[\theta_n^+ + T- \frac{\sigma}{\lambda}\rho_n^+ \left(1-\mathrm{e}^{-\lambda T}\right) \right] \!\!\!\! \mod 1  \label{eq:linear3mt} , \\
	\rho_{n+1} & = \rho_n^+ \mathrm{e}^{-\lambda T} \label{eq:linear4mt}\, .
\end{align}
Note that this has been constructed using a matched asymptotic expansion, using (\ref{deltaapprox}), and is valid in the
limit $\tau \rightarrow 0$.
For weak forcing, where $\varepsilon\ll1$, $P_{1,2}$ vary slowly through the kick and can be approximated by their values at $(\theta_n,\rho_n)$ so that to $O(\varepsilon^2)$
\begin{align}
	\theta_{n+1} &= \theta_n + T + \varepsilon P_1(\theta_n,\rho_n) + \frac{\sigma}{\lambda}\Big(\rho_n+\varepsilon P_2(\theta_n)\Big) \left(1-\e^{-\lambda T}\right), 
	\label{eq:2dmap1} \\
	\rho_{n+1} &= \left(  \rho_n + \varepsilon P_2(\theta_n) \right) \e^{-\lambda T}\, .
	\label{eq:2dmap2}
\end{align}
Although this explicit map is convenient for numerical simulations we prefer to work with the full stroboscopic map (\ref{eq:linear3mt})-(\ref{eq:linear4mt}), which is particularly useful for comparing and contrasting the behaviour of different planar single neuron models with arbitrary kick strength. As an indication of the presence of chaos in the dynamics resulting from this system, we evaluate the largest Lyapunov exponent
of the map (\ref{eq:linear3mt}-\ref{eq:linear4mt}) by numerically evolving a tangent vector and computing its rate of growth (or contraction); see e.g. \cite{Christiansen1997} for details.

In Fig.~\ref{fig:kicksMLFHN} we compare the functions $P_{1,2}$ for both the FHN and the ML models.  We note that $P_2$ for the FHN model is near $0$ for a large set of $\theta$, whilst the same is true for $P_1$ for the ML model. This means that kicks in the FHN model will tend to primarily cause phase shifts, whilst the same kicks in the ML model will primarily cause shifts in amplitude. 

\begin{figure}[h!]
\begin{center}
\includegraphics[width=6.in]{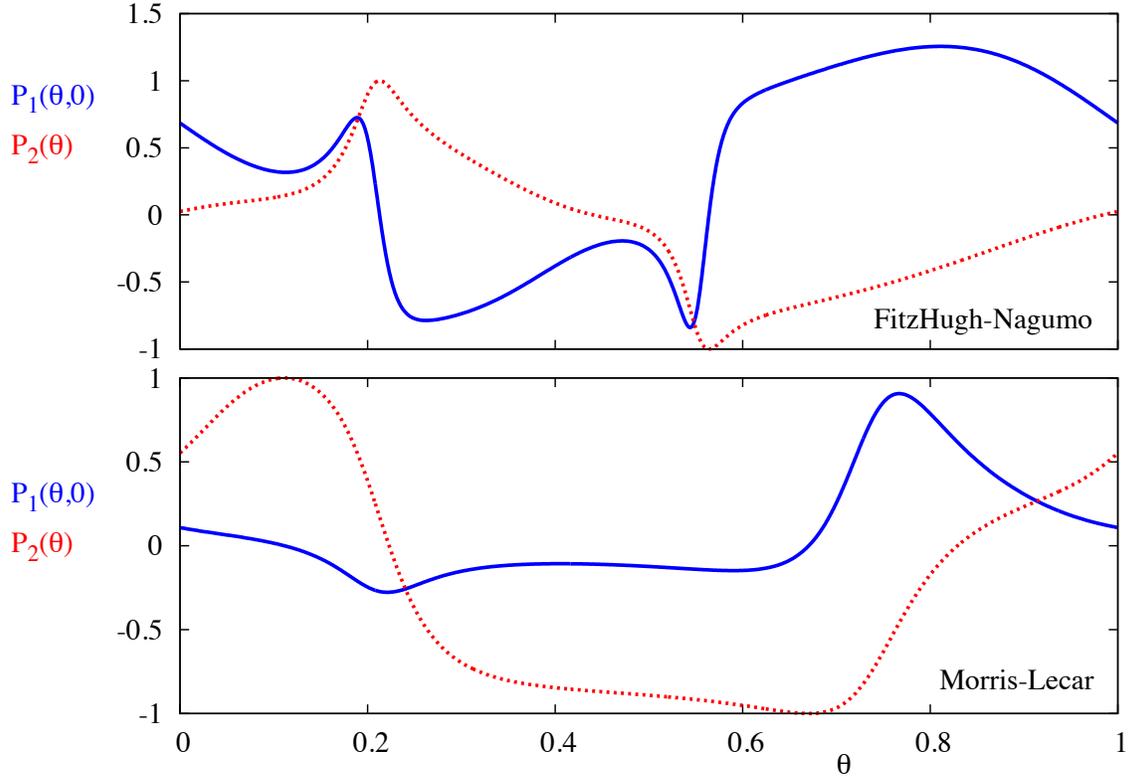} 
\caption{ \label{fig:kicksMLFHN} The blue curves show the change in $\theta$ under the action of a pulsatile kick in $v$, whilst the red dashed curves show the change in $\rho$ under the same kick. The top plot is for the FHN model, whilst the bottom plot is for the ML model. We evaluate the effect of the kicks at $\rho_n=0$, where we observe the largest changes in $\theta$ under the action of kicks.}
\end{center}
\end{figure}
\begin{figure}
\begin{center}
\includegraphics[width=6in]{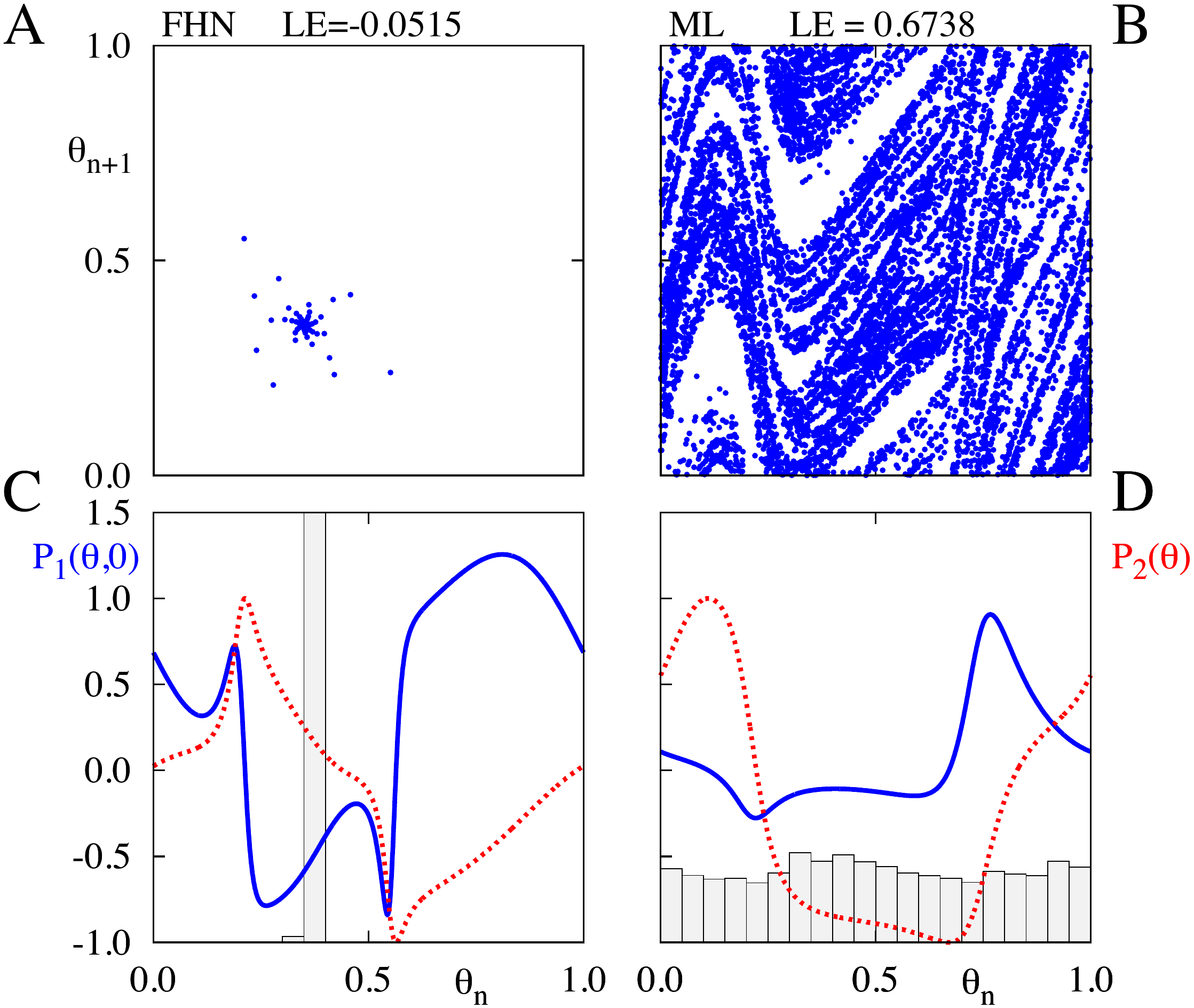}
\caption{ \label{fig:AllMapsAndHist} 
Panel A shows successive iterates of $\theta$ in system \eqref{eq:linear3mt}-\eqref{eq:linear4mt} with functions $P_{1,2}$ taken from the FHN model, whilst panel B presents the same iterates but for functions $P_{1,2}$ from the ML model. Panel C shows $P_{1,2}$ for the FHN model, where the bold blue line is $P_1$ and the red dashed line is $P_2$. Superimposed on this panel is a histogram displaying how kicks are distributed in terms of $\theta$ alone. Panel D shows the same information, except this time for forcing functions from the ML model. Parameter values are $\sigma=3.0$, $\lambda=0.1$, $\varepsilon=0.1$, and $T=2.0$.} 
\end{center}
\end{figure}

We plot in the top row of Fig.~\ref{fig:AllMapsAndHist} the pair $(\theta_n,\theta_{n+1})$, for 
\eqref{eq:2dmap1}-\eqref{eq:2dmap2} for the FHN and ML models. For the FHN model, we find that the system has Lyapunov exponent of $-0.0515 < 0$. For the ML model the Lyapunov exponent is $0.6738> 0$. This implies that differences in the functional forms of $P_{1,2}$ can help to explain the generation of chaos.

Now that we know the relative contribution of kicks in $v$ to kicks in $(\theta,\rho)$, it is also useful to know where kicks actually occur in terms of $\theta$ as this will determine the contribution of a train of kicks to the $(\theta,\rho)$ dynamics. In Figs.~\ref{fig:AllMapsAndHist} C and D we plot the distribution of kicks as a function of $\theta$. For the ML model we observe that the kicks are distributed over all phases, while for FHN model there is a grouping of kicks around the region where $P_2$ is roughly zero.
This means that kicks will not be felt as much in the $\rho$ variable, and so trajectories here do not get kicked far from cycle. This helps explain why it is more difficult to generate chaotic responses in the FHN model.

After transients, we observe a $1:1$ phase-locked state for the FHN model. For a phase-locked state, small perturbations will ultimately decay as the perturbed trajectories also end up at the phase-locked state after some transient behaviour. This results in a negative largest Lyapunov exponent of $-0.0515$. We note the sharply peaked distribution of kick phases, which is to be expected for discrete-time systems possessing a negative largest Lyapunov exponent, since such systems tend to have sinks in this case. The phase-locked state here occurs where $P_2$ is small, suggesting that trajectories stay close to the limit cycle. Since kicks do not move move trajectories away from cycle, there is no possibility of folding, and hence no chaotic behaviour. For the ML model, we observe chaotic dynamics around a strange attractor, where small perturbations can grow, leading to a positive largest Lyapunov exponent of $0.6738$. This time, the kicks are distributed fairly uniformly across $\theta$, and so, some kicks will take trajectories away from the limit cycle, thus leading to shear-induced folding and chaotic behaviour. 

\section{Discussion}
\label{sec:discussion}

In this paper, we have used the notion of a moving orthonormal coordinate system around a limit cycle to study dynamics in a neighbourhood around it. This phase-amplitude coordinate system can be constructed for any given ODE system supporting a limit cycle. A clear advantage of the transformed description over the original one is that it allows us to gain insight into the effect of time dependent perturbations, using the notion of shear, as we have illustrated by performing case studies of popular neural models, in two and higher dimensions.
Whilst this coordinate transformation does not result in any reduction in dimensionality in the system, as is the case with classical phase reduction techniques, it opens up avenues for moving away from the weak coupling limit, where $\varepsilon \rightarrow 0$. Importantly, it emphasises the role of the two functions $P_1(\theta,\rho)$ and $P_2(\theta)$ that provide more information about inputs to the system than the iPRC alone. It has been demonstrated that moderately small perturbations can exert remarkable influence on dynamics in the presence of other invariant structures \cite{Lin2012}, which can not be captured by a phase only description. In addition, small perturbations can accumulate if the timescale of the perturbation is shorter than the timescale of attraction back to the limit cycle. This should be given particular consideration in the analysis of neural systems, where oscillators may be connected to thousands of other units, so that small inputs can quickly accumulate.

One natural extension of this work is to move beyond the theory of weakly coupled oscillators to develop a framework for describing neural systems as networks of phase-amplitude units.  This has previously been considered for the case of weakly coupled weakly dissipative networks of nonlinear planar oscillators (modelled by small dissipative perturbations of a Hamiltonian oscillator) \cite{Ashwin1989,Ashwin2000a,Ashwin2005}
.  It would be interesting to develop these ideas and obtain network descriptions of the following type
\begin{align}
\dot{\theta}_i &= 1 + f_1(\theta_i,\rho_i) +\sum_j w_{ij} H_1(\theta_i, \theta_j,\rho_i, \rho_j) ,\\
\dot{\rho}_i &= A(\theta_i)\rho_i + \sum_j w_{ij} H_2(\theta_i, \theta_j,\rho_i, \rho_j),
\end{align}
with an appropriate identification of the interaction functions $H_{1,2}$ in terms of the biological interaction between neurons and the singe neuron functions $P_{1,2}$.  Such phase-amplitude network models are ideally suited to describing the behaviour of the mean-field signal in networks of strongly gap junction coupled ML neurons \cite{Han95,Coombes2008}, which is known to vary because individual neurons make transitions between cycles of different amplitudes. 
Moreover, in the same network weakly coupled oscillator theory fails to explain how the synchronous state can stabilise with increasing coupling strength (predicting that it is always unstable), as observed numerically.  
All of the above are topics of ongoing research and will be reported upon elsewhere.


\appendix

\section*{Appendices}

\section{Derivation of the transformed dynamical system}
\label{app:trans}
Starting from 
\begin{equation}
	\dot{x} = f(x)+\epsilon g(x,t) ,
\end{equation}
we make the transformation $x(t)=u(\theta(t))+\zeta(\theta(t))\rho(t)$, giving
\begin{equation}
	\left[\FD{u(\theta)}{\theta}+\FD{\zeta(\theta)}{\theta}\rho\right]\dot{\theta}+\zeta(\theta)\dot{\rho}=f(u(\theta)+\zeta(\theta)\rho)+\epsilon g(u(\theta+\zeta(\theta)\rho,t).
\label{eq:1.5}
\end{equation}
We proceed by projecting \eqref{eq:1.5} onto $\xi(\theta)$, using (\ref{xi}).
The left hand side of \eqref{eq:1.5} now reads:
\begin{equation}
	\left[\left|\FD{u}{\theta}\right|+\xi^T\FD{\zeta}{\theta}\rho\right]\FD{\theta}{t},
\end{equation}

\noindent where $\xi^T$ denotes the transpose of $\xi$ and the right hand side of \eqref{eq:1.5} becomes
\begin{align}
	&\xi^Tf(u+\zeta\rho)+\epsilon \xi^Tg(u+\zeta\rho) \nonumber \\
	&=\left[\left|\FD{u}{\theta}\right|+\xi^T\FD{\zeta}{\theta}\rho\right]+\xi^Tf(u+\zeta\rho)-\xi^Tf(u)-\xi^T\FD{\zeta}{\theta}\rho +\epsilon \xi^Tg(u+\zeta\rho,t).
\end{align}
Thus,
\begin{align}
	\dot{\theta}=1+f_1(\theta,\rho)+\epsilon h^T(\theta,\rho)g(u(\theta)+\zeta(\theta,t)\rho,t),
	\label{dthetadt}
\end{align}
where
\begin{equation}
	h(\theta,\rho)=\left[\left|\FD{u}{\theta}\right|+\xi^T\FD{\zeta}{\theta}\rho\right]^{-1}\xi(\theta) ,
\end{equation}
and
\begin{equation}
	f_1(\theta,\rho)=-h^T(\theta,\rho)\FD{\zeta}{\theta}\rho(\theta)+h^T(\theta,\rho)\left[f(u+\zeta\rho)-f(u)\right].
\end{equation}

\noindent Upon projecting both sides of \eqref{eq:1.5} onto $\zeta(\theta)$, the left hand side reads
\begin{align}
	&\zeta^T\left[\FD{u}{\theta}+\FD{\zeta}{\theta}\right]\FD{\theta}{t}+\FD{\rho}{t} =\zeta^T\FD{\zeta}{\theta}\rho\FD{\theta}{t}+\FD{\rho}{t} 
	=\zeta^T\FD{\zeta}{\theta}\rho\left[1+f_1(\theta,\rho)+\epsilon h^T(\theta,\rho)g(u+\zeta\rho,t)\right]+\FD{\rho}{t},
\end{align}
whilst the right hand side becomes
\begin{align}
	\zeta^Tf(u+\zeta\rho)+\epsilon\zeta^T g(u+\zeta\rho,t) 
	=-\zeta^T f(u)+\zeta^T \Df \zeta-\zeta^T \Df \zeta+\zeta^Tf(u+\zeta\rho)+\epsilon g(u+\zeta\rho,t),
\end{align}
since $\zeta^Tf(u)=\zeta^T \d u/\d\theta=0$ and where $\Df$ denotes the Jacobian of $f$. Putting together the previous two equations yields
\begin{align}
	\dot{\rho}=A(\theta)\rho+f_2(\theta,\rho)+\epsilon\zeta^T\left[\mathrm{I}-\FD{\zeta(\theta)}{\theta}\rho h(\theta,\rho)\right]g(u(\theta)+\zeta(\theta)\rho,t),
\label{drhodt}
\end{align}
where
\begin{align}
	A(\theta)=\zeta^T\left[-\FD{\zeta}{\theta}+\Df\zeta\right],
\end{align}
\begin{align}
	f_2(\theta,\rho)=-\zeta^T\FD{\zeta}{\theta}\rho f_1+\zeta^T\left[f(u+\zeta\rho)-f(u)-\Df\zeta\rho\right] .
\end{align}
It may be easily seen that $f_1(\theta,\rho)=O(\rho)$ as $\rho\rightarrow0$ and that $f_2(\theta,0)=0$ and $\partial f_2(\theta,0)/\partial\rho=0$.
\noindent Overall, combining \eqref{dthetadt} and \eqref{drhodt} we arrive at the transformed system:
\begin{align}
	\dot{\theta}&=1+f_1(\theta,\rho)+\epsilon h^T(\theta,\rho)g(u(\theta)+\zeta(\theta,t)\rho,t), \nonumber \\
	\dot{\rho}&=A(\theta)\rho+f_2(\theta,\rho)+\epsilon\zeta^T\left[\mathrm{I}-\frac{d\zeta(\theta)}{d\theta}\rho h(\theta,\rho)\right]g(u(\theta)+\zeta(\theta)\rho,t).
\end{align}

In order to evaluate the functions $f_1,f_2$ and $A$ for models with dimension larger than two, we need to calculate $\d \zeta / \d\theta$. Defining by  $\gamma_i(\theta)$, the direction angles of $\xi(\theta)$, we have that 
\begin{align}
	\zeta_i = \e_i - \frac{\cos\gamma_i}{1+\cos \gamma_1}\left(\e_1+\xi(\theta)\right) = \e_i - \frac{\e_i\cdot \xi(\theta)}{1+\e_1\cdot \xi(\theta)}\left(\e_1+\xi(\theta)\right) , \quad i=2,3,\dots, n ,
\end{align}
where the index $i$ denotes the column entry of $\zeta$ and $x\cdot y$ denotes the dot product between vectors $x$ and $y$. Defining
\begin{equation}
	u_i(\theta) = \frac{\e_i\cdot \xi(\theta)}{1+\e_1\cdot \xi(\theta)},
\end{equation}
and
\begin{equation}
	w_j(\theta) = 1+\e_{1,j}\cdot \xi_j(\theta),
\end{equation}
where $j$ denotes the row index, we have
\begin{align}
	\FD{\zeta_{i,j}}{\theta} &= -u\FD{w_j}{\theta}-w_j\FD{u_i}{\theta}.
\end{align}
By the quotient rule for vectors we find that,
\begin{equation}
	\FD{u_i}{\theta} = \frac{\left(1+\e_1\right)\left(\e_i\cdot\FD{\xi}{\theta}\right)-\left(\e_i\cdot \xi(\theta)\right)\left(\e_1\FD{\xi}{\theta}\right)}{\left(1+\e_1\cdot \xi(\theta) \right)^2},
\end{equation}
and that
\begin{equation}
	\FD{w_j}{\theta} = \FD{\xi_j}{\theta}.
\end{equation}
Overall, we have that
\begin{equation}
	\FD{\zeta_{i,j}}{\theta} = - \frac{\e_i\cdot \xi(\theta)}{1+\e_1\cdot \xi(\theta)} - \left(\e_{1,j}+\xi_j(\theta)\right)\left(\frac{\left(1+\e_1\right)\left(\e_i\cdot\FD{\xi}{\theta}\right)-\left(\e_i	\cdot \xi(\theta)\right)\left(\e_1\FD{\xi}{\theta}\right)}{\left(1+\e_1\cdot \xi(\theta) \right)^2}\right).
\end{equation}

\section{Gallery of models}
\label{sec:models}

\subsection{Morris-Lecar}
\label{ssec:ML}
The ML equations describe the interaction of membrane voltage with just two ionic currents: Ca$^{2+}$ and $K^+$.  Membrane ion channels are selective
for specific types of ions; their dynamics are modelled here by the gating variable $w$ and the auxiliary functions $w_\infty$, $\tau_w$, and $m_\infty$.  The latter have the form
\begin{align}
   m_\infty(v) &= \frac12\big[1+\tanh\big((v-v_1)/v_2\big)\big], \qquad 
   \tau_w(v) = 1/\cosh\big((v-v_3)/(2v_4)\big), \nonumber \\
   w_\infty(v) &= \frac12\big[1+\tanh\big((v-v_3)/v_4\big)\big].
 \end{align}

The function $m_\infty(v)$ models the action of fast voltage-gated calcium ion channels; $v_{\rm Ca}$ is the reversal (bias) potential for the calcium current and $g_{\rm Ca}$ the corresponding conductance. The functions $\tau_w(v)$ and $w_\infty(v)$ similarly describe the dynamics of slower-acting potassium channels, with its own reversal potential $v_{\rm K}$ and conductance $g_{\rm K}$.  The constants $v_{\rm leak}$ and $g_{\rm leak}$ characterize the leakage current that is present even when the neuron is in a quiescent state.
Parameter values are $C=20.0 \, \mu\text{F}/\text{cm}^2, \; g_\text{l}=2.0 \,\text{mmho}/\text{cm}^2, \; g_\text{k}=8.0 \,\text{mmho}/\text{cm}^2, \; g_\text{ca}=4.0 \,\text{mmho}/\text{cm}^2, \; \phi=0.23, \; I=39.5 \,\mu\text{A}/\text{cm}^2, \; v_\text{l}=-60.0 \,\text{mV}, \; v_\text{k}=-84.0  \,\text{mV}, \; v_\text{ca}=120.0  \,\text{mV}, \; v_1=-1.2  \,\text{mV},\: v_2=18.0  \,\text{mV}, \; v_3=12.0  \,\text{mV}$, and $v_4=17.4 \,\text{mV}$.

\subsection{Reduced Connor-Stevens Model}
\label{ssec:CS}

For the reduced CS model, we start with the full Hodgkin-Huxley model, with $m,n,h$ as gating variables and use the method of equivalent potentials as treated in \cite{Kepler1992}, giving rise to the following form for the function $g$
\begin{equation}
	G(v,u) = \left( \PD{F}{h} \left [ \frac{h_\infty(v) - h_\infty
	(u)}{\tau_h(v)} \right ] + \PD{F}{n} \left [ \frac{n_\infty(v) -
	n_\infty(u)}{\tau_n(v)} \right ] \right) \Big/ \left( \PD{f}{h_\infty}
\FD{h_\infty(u)}{u} + \PD{f}{n_\infty} \FD{n_\infty(u)}{u} \right)
\end{equation}
where $\partial F /\partial h$ and $\partial F /\partial n$ are
evaluated at $h = h_\infty(u)$ and $n = n_\infty(u)$.  
For the gating variables $(a,b)$, we have
\begin{align}
	a_\infty(v) &= \left(\frac{0.0761{\rm e}^{\frac{v+94.22}{31.84}}}{1+{\rm e}^{\frac{v+1.17}{28.93}}}\right)^\frac{1}{3} , & \tau_a(v) &= 0.3632+\frac{1.158}{1+{\rm e}^{\frac{v+55.96}{20.12}}} , \\
	b_\infty(v) &= \left(\frac{1}{1+{\rm e}^{\frac{v+53.3}{14.54}}}\right)^4, & \tau_b(v) &= 1.24+\frac{2.678}{1+{\rm e}^{\frac{v+50}{16.027}}} .
\end{align}
\textcolor{red}{Parameter values are $C=1 \,\mu\text{F}/\text{cm}^2, \: g_\text{L}=0.3 \,\text{mmho}/\text{cm}^2,\: g_\text{K}=36.0 \,\text{mmho}/\text{cm}^2, \: g_\text{a}=47.7 \,\text{mmho}/\text{cm}^2, \; I=35.0 \,\mu\text{A}/\text{cm}^2,\; v_0=80.0 \,\text{mV},\;v_\text{a}=-75.0 \,\text{mV},\; v_\text{K}=-77.0 \,\text{mV},\;v_\text{L}=-54.4 \,\text{mV}$, and $v_\text{Na}=50.0 \,\text{mV}$.}

\subsection{FitzHugh-Nagumo Model}
\label{ssec:FHN}

The FHN model is a phenomenological model of spike generation, comprising of 2 variables. The first, represents the membrane potential and includes a cubic nonlinearity, whilst the second variable is a gating variable, similar to $w$ in the ML model, which may be thought of as a recovery variable. The system is
  \begin{equation}
  	\mu \dot{v} = v(a-v)(v-1) + I - w , \qquad 
	\dot{w} = v - bw , \label{eq:FHN}
  \end{equation}
where we use the following parameter values:  $\mu=0.05$, $a=0.9$, $I=1.1$, and $b=0.5$.


{\ifthenelse{\boolean{publ}}{\footnotesize}{\small}
 \bibliographystyle{bmc_article}  
  \bibliography{PhaseAmplitude} }     



\ifthenelse{\boolean{publ}}{\end{multicols}}{}



\clearpage

\section*{List of abbreviations}
\begin{itemize}
	\item ML - Morris-Lecar
	\item FHN - FitzHugh-Nagumo
	\item CS - Connor-Stevens
	\item LE - Lyapunov exponent
\end{itemize}

\section*{Additional files}
The additional files for this paper may be found on the additional files page at \url{http://www.maths.nottingham.ac.uk/personal/pmxkw2/MultiplePages/Welcome.html}

\end{bmcformat}
\end{document}